\documentclass[preprint,
amsmath,amssymb,
aps,superscriptaddress
]{revtex4-1}
\usepackage{dcolumn}
\usepackage{bm}
\usepackage{caption}
\usepackage{mathrsfs}
\usepackage[titletoc]{appendix}
\usepackage{graphicx}
\usepackage{epstopdf}
\usepackage{hyperref}
\usepackage{subfigure}
\usepackage{slashed}
\usepackage{tikz}
\usepackage{diagbox}
\allowdisplaybreaks[4]
\begin{document}
	
	\title{Distinguishing Quantum Matter by Gravity with Differential Scattering Cross Section at Tree Level}
	
	\author{Xue-Nan Chen}
	\email{For correspondence: xnchen@hust.edu.cn}
	\affiliation{Group of Fundanmental Physics, Hubei Key Laboratory of Gravitational and Quantum Physics, School of Physics, Huazhong University of Science and Technology, Wuhan 430074, China}

	\begin{abstract}
	The definition of weak equivalence principle of quantum matter is an open problem at present. In order to reflect the probability of quantum system in the quantum version of weak equivalence principle, we proposed a quantum weak equivalence principle based on differential scattering cross section at tree level, that is, the differential scattering cross section does not depend on the mass and properties of the scattered particles when the target particles take the large mass limit. This version of the quantum equivalence principle we proposed will be broken by the spin properties of quantum matter. In the non-relativistic case, the difference of differential scattering cross sections of scattered particles with different spin properties scattered by target particles is mainly reflected in the order of $ \mathcal O (p _ {\mathrm{cm}} ^2) $. In the relativistic case , we studied the asymptotic behavior of differential scattering cross sections at small angles. When the target particles are scalar particles, the difference of light particles with different spin properties is mainly reflected in the $ \mathcal O (1/\theta^2) $ order. When the target particles are Dirac particles, the difference of light particles with different spin properties is mainly reflected in the $ \mathcal O (1/\theta^4) $ order.  The polarization of differential scattering cross section when scattered particles are Dirac particles is investigated. The result of the degree of polarization depends on the polarization direction of the incident particles.
	\end{abstract}

	\maketitle
	\section{Introduction}
	
	The classical weak equivalence principle (WEP) requires that the trajectories of classical point particles with specified initial positions and velocities are identical, and the trajectories do not depend on the mass and other properties of classical point particles. Whether quantum particles will still maintain the WEP has attracted many theoretical and experimental research \cite{Colella75,Alvarez97,Viola97,Peters99,Fray04,Chowdhury12,Oriando16,Rosi17,Schwartz19,Flores19}.
	Compared with classical particles, quantum particles carry other properties, such as matter/antimatter \cite{Charman13,Hohensee13,Hamilton14}, spin \cite{Ni10,Tarallo14,Duan16} and internal structure \cite{Fray04,Rosi17,Schlippert14}, which will challenge the WEP. At the same time, the test of WEP of quantum system will help us understand how quantum matter interacts with gravity. At present, almost all theories that try to unify the standard model of gravity theory and particle physics require breaking the WEP \cite{Damour12,Lammerzahl03}.
	
	Although the investigations on the WEP of quantum system have a wide range, from the effects described by Schr\"{o}dinger equation with gravitational potential to the effects originated from the quantized gravitational field \cite{Davies82,Dalvt99,Ali06,Accioly08,Mousavi15,Tino21}, there is controversy about whether the weak equivalence principle of quantum system is broken in theory. This is due to the fact that defining the WEP in its quantum version is still a very open topic \cite{Herdegen02,Okon11,Zych18}. Because quantum systems have characteristics different from classical point particles, such as nonlocality and probability, we cannot directly translate the classical weak equivalence principle into the corresponding quantum version.
	
	One of the attempts is to construct some indirect observables from the statistical meaning of physical observables, such as the expected value of physical operators and their functions \cite{Anastopoulos18,Wang24}, Fisher information \cite{Seveso17}, the trajectories defined by multipole expansion of the energy-momentum tensor(EMT) (the Mathisson-Papapetrou-Dixon (MPD) equations) \cite{Mathisson37,Papaetrou51,Dixon70a,Dixon70b}, etc. However, the trajectories given by MPD equation is not unique, and it needs additional constraints to be determined. Different constraints give different trajectory results \cite{Corinaldesi51,Pirani56,Duval19}, and these constraints and trajectories are not well related to the actual experiment. Lian et al. proposed to solve the multipole expansion of the EMT by using wave packet evolution, and obtained the centre of the EMT \cite{Lian22}. Using this method, they predicted the birefringence of electromagnetic wave packets given by different components of the EMT and different types of the EMT in the gravitational field. 
	
	Another attempt is to use direct observables, such as probability amplitude \cite{Rosi17, Duan16,Schlippert14},  probability distribution \cite{Ali06,Anastopoulos18},  etc., or to construct some indirect observables based on probability distribution, such as arrival time distribution \cite{Mousavi15}, etc. At present, the common method is to extract the phase information of the probability amplitude in the atomic interferometer, get the atomic gravity acceleration from it, and define the E\"{o}tv\"{o}s ratio for the gravity acceleration of different atoms \cite{Schlippert14}. However, the method of probability amplitude is based on non-relativistic quantum mechanics, which is obviously not suitable for high-energy particles, and the probability amplitude lacks the classical correspondence, so it is difficult to describe the WEP uniformly between quantum and classical.
	
	The differential scattering cross section  not only can be obtained by using the classical mechanics method at the classical level, but also  can reflect the probability transition process of quantum states at the quantum level. Therefore, using the differential scattering cross section as a pointer to test the WEP of quantum system can not only reflect the probability process of quantum mechanics well, but also correspond to the classical process well.
	
	At present, the research of scattering amplitude used to test WEP of quantum system can be roughly divided into two categories. One treats the external gravitational field as an external source to calculate the differential scattering cross sections of particles with different spin properties under the external source \cite{Accioly08,Accioly09}. The other considers the scattering of two bodies, takes the mass of one particle as infinity, and calculates the bending angle from the scattering amplitude \cite{BB15,BB16,Bai17,Chi19}. The second-order effect is calculated in these two kinds of studies, and it is found that the next-to-leading-order effect depends on the mass or energy of particles, which means the WEP of quantum system is broken. For the former, only the scattering of particles with different spins scattered by massive scalar particles with infinite mass is investigated. For the latter, the current research is only limited to the non-rotating mass object as an external source and the scattered particles are massless. Because the massive particles are not be considered, the broken of quantum WEP at tree level is not discovered,  and the bending angle is indirect observable for quantum matter. 
	
	Based on the perturbation gravity theory, we will calculate the differential scattering cross sections between particles with different spin properties interacting with each other by gravitational interaction, and take the mass of one particle as infinity to study the behavior of its differential scattering cross sections in non-relativistic and relativistic cases. Our paper is organized as follows.  In Sect. \ref{sect:2}, we will explain the definition of quantum weak equivalence principle under differential scattering cross section. In Sect. \ref{sect:3}, we will calculate the differential scattering cross sections of two scalar particles, a scalar particle and a Dirac particle, and two Dirac particles separately, and analyze its behavior in non-relativistic and relativistic cases, taking the mass of a particle as infinite. In Sect. \ref{sect:4}, we will calculate the degree of polarization of the differential scattering cross section $P$ with Dirac particles, and point out that the degree of polarization of the differential scattering cross section depends on the initial polarization direction. In Sect. \ref{sect:5}, we will summarize our results and discuss their physical implications.
	
	\section{The differential scattering cross section and Quantum WEP}
	\label{sect:2}
	
At the beginning, we introduce the concept of quantum WEP through the differential scattering cross section:
	
	\textit{When two scattered particles with different masses or different properties (such as spin properties) with the same initial velocity are scattered by a target particle, the results of differential scattering cross section are the same when the mass of the target particle takes the large mass limit.}
	
	Here we need to pay attention to the following points:
	 
	 1. The differential scattering cross section here refers to the result of not distinguishing the final polarization state of polarized particles and averaging the initial polarization state of polarized particles. (Distinguishing the polarization state of scattered particles can also define a new observable, that is, the degree of polarization. We will discuss it in Sect. \ref{sect:4}.)
	 
	 2. The target particles here need to take a large mass limit in order to avoid the recoil effect. The recoil effect makes it impossible to test the equivalence principle in the same gravitational background.
	 
	 3. In order to test the quantum WEP, the spin properties of target particles should be the same to ensure that scattered particles with different spin properties are in the same gravitational field background.
	 
	 The unpolarized differential scattering cross section is defined as:
	 \begin{equation}{\label{1}}
	 	\begin{aligned}
	 		\frac{\mathrm{d} \sigma}{\mathrm{d }\Omega}= & \frac{\sqrt{m_1^2+p_{\mathrm{cm}}^2} \sqrt{m_2^2+p_{\mathrm{cm}}^2}}{\sqrt{\left(p_{\mathrm{cm}}^2+\sqrt{m_1^2+p_{\mathrm{cm}}^2} \sqrt{m_2^2+p_{\mathrm{cm}}^2}\right)^2-m_1^2 m_2^2}} \\
	 		& \times \frac{p_{\mathrm{cm}}}{64 \pi^2 \sqrt{m_1^2+p_{\mathrm{cm}}^2} \sqrt{m_2^2+p_{\mathrm{cm}}^2}\left(\sqrt{m_1^2+p_{\mathrm{cm}}^2}+\sqrt{m_2^2+p_{\mathrm{cm}}^2}\right)}|M|^2,
	 	\end{aligned}
	 \end{equation}
 where,
 \begin{equation}{\label{2}}
 	|M|^2=\frac{1}{n_mn_M} \sum_{s_m,s_M} \sum_{s_m^{\prime}, s_M^{\prime}}\left|M_{s_m, s_M, s_m^{\prime}, s_M^{\prime}}\right|^2.
 \end{equation}
Hence, the mathematical expression of the quantum equivalent principle under the definition of differential scattering cross section is:
\begin{equation}
	\frac{\mathrm{d} \sigma}{\mathrm{d} \Omega}(m_1,S_1,M\rightarrow\infty)=\frac{\mathrm{d} \sigma}{\mathrm{d} \Omega}(m_2,S_2,M\rightarrow\infty),
\end{equation}
where, $S$ represents the spin property. For scalar particle, $ S=0$ and for Dirac particle, $S=1/2$. The types of differential scattering cross sections calculated in this paper are shown as TAB. \ref{tab:1}. And we will analyze the asymptotic behavior of the results of differential scattering cross section in the following two cases: 1)non-relativistic limit ($p_{\mathrm{cm}}\rightarrow 0$); 2)the small angle limit($\theta\rightarrow 0$) in relativistic limit.
\begin{table}[htbp]
	\centering
	\caption{The types of differential scattering cross sections calculated in this paper}
\begin{tabular}{|c|c|c|}
	\hline
	\diagbox{scattered particle}{target particle}& scalar & Dirac\\
	\hline
	scalar&$\mathrm{d}\sigma_{ss}/\mathrm{d}\Omega$ &$\mathrm{d}\sigma_{sd}/\mathrm{d}\Omega(m_s\ll m_d)$ \\
	\hline
	Dirac&$\mathrm{d}\sigma_{sd}/\mathrm{d}\Omega(m_d\ll m_s)$  &$\mathrm{d}\sigma_{dd}/\mathrm{d}\Omega$ \\
	\hline
\end{tabular}
\label{tab:1}
\end{table}

	\section{The differential scattering cross section of two particles}
	\label{sect:3}

In this paper, we choose the perturbative quantum gravity, which is a quantum theory describing small metric perturbations $h_{\mu\nu}$, associated with gravitons, that propagate on a flat background spacetime $\eta_{\mu\nu}$. Here, $\eta_{\mu\nu}=\mathrm{diag}(1,-1,-1,-1)$. The complete spacetime metric $g_{\mu\nu}$ accounts both for the background and for 	the perturbations:
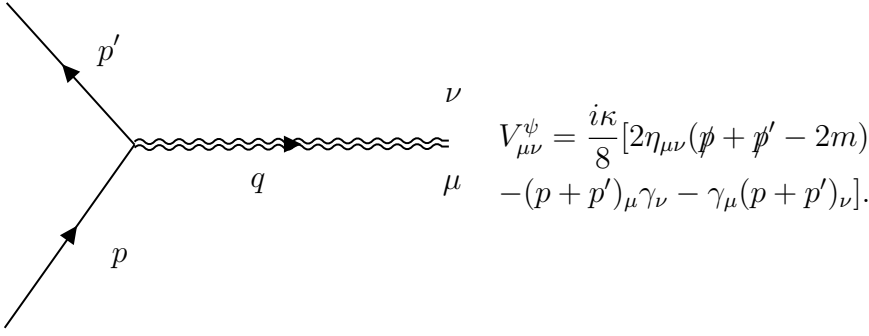
\begin{equation}
	g_{\mu\nu}=\eta_{\mu\nu}+\kappa h_{\mu\nu},
\end{equation}
where, $\kappa=\sqrt{32\pi G}$ is the gravitational coupling expressed via the Newton constant $G$. Then we can obtain the infinite expansions for $g^{\mu\nu}$, $\sqrt{-g}$,  Christoffel symbols, and Riemann tensor. Consequently, the microscopic action of general relativity  presents an infinite perturbative expansion containing infinite interaction terms. We will expand the matter field section in the same way. Then we obtain the action:
\begin{equation}
	S=\int\mathrm{d}^4x (\mathscr{L}_G^{(0)}+\mathscr{L}_{\text{mat}}^{(0)}-\frac\kappa2 h_{\mu\nu}T^{\mu\nu(0)}_{\text{mat}})+\cdots,
\end{equation}
where,
\begin{equation}
	\mathscr{L}_G^{(0)}=-\frac{1}{2} h^{, \nu} h_{, \nu}+\frac{1}{2} h_{\mu \nu}^{, \lambda} h_{, \lambda}^{\mu \nu}+h_{, \nu} h_{, \lambda}^{\nu \lambda}-h_{, \mu}^{\nu \lambda} h_{\mu \nu}^{, \lambda},
\end{equation}
and $\mathscr{L}^{(0)}_{\text{mat}}$ and $T^{\mu\nu(0)}_{\text{mat}}$ is Lagrangian and energy-momentum tensor in the flat background. We can derive the Feynman rules for the gravitational propagator (in the de Donder gauge) and interaction vertices directly from the action \cite{Choi95}:
\tikzset{every picture/.style={line width=0.75pt}} 

\begin{tikzpicture}[x=0.75pt,y=0.75pt,yscale=-1,xscale=1]
	
	\draw    (104.01,119.5) .. controls (105.69,117.85) and (107.36,117.86) .. (109.01,119.53) .. controls (110.66,121.21) and (112.33,121.22) .. (114.01,119.57) .. controls (115.68,117.92) and (117.35,117.93) .. (119.01,119.6) .. controls (120.67,121.27) and (122.34,121.28) .. (124.01,119.63) .. controls (125.69,117.98) and (127.36,117.99) .. (129.01,119.67) .. controls (130.67,121.34) and (132.34,121.35) .. (134.01,119.7) .. controls (135.69,118.05) and (137.36,118.06) .. (139.01,119.74) .. controls (140.67,121.41) and (142.34,121.42) .. (144.01,119.77) .. controls (145.68,118.12) and (147.35,118.13) .. (149.01,119.8) .. controls (150.66,121.48) and (152.33,121.49) .. (154.01,119.84) .. controls (155.68,118.19) and (157.35,118.2) .. (159.01,119.87) .. controls (160.67,121.54) and (162.34,121.55) .. (164.01,119.9) .. controls (165.69,118.25) and (167.36,118.26) .. (169.01,119.94) .. controls (170.67,121.61) and (172.34,121.62) .. (174.01,119.97) .. controls (175.68,118.32) and (177.35,118.33) .. (179.01,120) .. controls (180.66,121.68) and (182.33,121.69) .. (184.01,120.04) .. controls (185.68,118.39) and (187.35,118.4) .. (189.01,120.07) .. controls (190.66,121.75) and (192.33,121.76) .. (194.01,120.11) .. controls (195.68,118.46) and (197.35,118.47) .. (199.01,120.14) .. controls (200.67,121.81) and (202.34,121.82) .. (204.01,120.17) .. controls (205.69,118.52) and (207.36,118.53) .. (209.01,120.21) .. controls (210.67,121.88) and (212.34,121.89) .. (214.01,120.24) .. controls (215.68,118.59) and (217.35,118.6) .. (219.01,120.27) .. controls (220.66,121.95) and (222.33,121.96) .. (224.01,120.31) .. controls (225.68,118.66) and (227.35,118.67) .. (229.01,120.34) .. controls (230.66,122.02) and (232.33,122.03) .. (234.01,120.38) .. controls (235.68,118.73) and (237.35,118.74) .. (239.01,120.41) .. controls (240.67,122.08) and (242.34,122.09) .. (244.01,120.44) .. controls (245.69,118.79) and (247.36,118.8) .. (249.01,120.48) .. controls (250.67,122.15) and (252.34,122.16) .. (254.01,120.51) .. controls (255.68,118.86) and (257.35,118.87) .. (259.01,120.54) .. controls (260.66,122.22) and (262.33,122.23) .. (264.01,120.58) .. controls (265.68,118.93) and (267.35,118.94) .. (269.01,120.61) .. controls (270.67,122.28) and (272.34,122.29) .. (274.01,120.64) .. controls (275.69,118.99) and (277.36,119) .. (279.01,120.68) .. controls (280.67,122.35) and (282.34,122.36) .. (284.01,120.71) .. controls (285.69,119.06) and (287.36,119.07) .. (289.01,120.75) .. controls (290.67,122.42) and (292.34,122.43) .. (294.01,120.78) .. controls (295.68,119.13) and (297.35,119.14) .. (299.01,120.81) .. controls (300.66,122.49) and (302.33,122.5) .. (304.01,120.85) .. controls (305.68,119.2) and (307.35,119.21) .. (309.01,120.88) .. controls (310.67,122.55) and (312.34,122.56) .. (314.01,120.91) -- (317.51,120.94) -- (317.51,120.94)(103.99,122.5) .. controls (105.66,120.85) and (107.33,120.86) .. (108.99,122.53) .. controls (110.64,124.21) and (112.31,124.22) .. (113.99,122.57) .. controls (115.66,120.92) and (117.33,120.93) .. (118.99,122.6) .. controls (120.65,124.27) and (122.32,124.28) .. (123.99,122.63) .. controls (125.67,120.98) and (127.34,120.99) .. (128.99,122.67) .. controls (130.65,124.34) and (132.32,124.35) .. (133.99,122.7) .. controls (135.67,121.05) and (137.34,121.06) .. (138.99,122.74) .. controls (140.65,124.41) and (142.32,124.42) .. (143.99,122.77) .. controls (145.66,121.12) and (147.33,121.13) .. (148.99,122.8) .. controls (150.64,124.48) and (152.31,124.49) .. (153.99,122.84) .. controls (155.66,121.19) and (157.33,121.2) .. (158.99,122.87) .. controls (160.65,124.54) and (162.32,124.55) .. (163.99,122.9) .. controls (165.67,121.25) and (167.34,121.26) .. (168.99,122.94) .. controls (170.65,124.61) and (172.32,124.62) .. (173.99,122.97) .. controls (175.66,121.32) and (177.33,121.33) .. (178.99,123) .. controls (180.64,124.68) and (182.31,124.69) .. (183.99,123.04) .. controls (185.66,121.39) and (187.33,121.4) .. (188.99,123.07) .. controls (190.64,124.75) and (192.31,124.76) .. (193.99,123.11) .. controls (195.66,121.46) and (197.33,121.47) .. (198.99,123.14) .. controls (200.65,124.81) and (202.32,124.82) .. (203.99,123.17) .. controls (205.67,121.52) and (207.34,121.53) .. (208.99,123.21) .. controls (210.65,124.88) and (212.32,124.89) .. (213.99,123.24) .. controls (215.66,121.59) and (217.33,121.6) .. (218.99,123.27) .. controls (220.64,124.95) and (222.31,124.96) .. (223.99,123.31) .. controls (225.66,121.66) and (227.33,121.67) .. (228.99,123.34) .. controls (230.64,125.02) and (232.31,125.03) .. (233.99,123.38) .. controls (235.66,121.73) and (237.33,121.74) .. (238.99,123.41) .. controls (240.65,125.08) and (242.32,125.09) .. (243.99,123.44) .. controls (245.67,121.79) and (247.34,121.8) .. (248.99,123.48) .. controls (250.65,125.15) and (252.32,125.16) .. (253.99,123.51) .. controls (255.66,121.86) and (257.33,121.87) .. (258.99,123.54) .. controls (260.64,125.22) and (262.31,125.23) .. (263.99,123.58) .. controls (265.66,121.93) and (267.33,121.94) .. (268.99,123.61) .. controls (270.65,125.28) and (272.32,125.29) .. (273.99,123.64) .. controls (275.67,121.99) and (277.34,122) .. (278.99,123.68) .. controls (280.65,125.35) and (282.32,125.36) .. (283.99,123.71) .. controls (285.67,122.06) and (287.34,122.07) .. (288.99,123.75) .. controls (290.65,125.42) and (292.32,125.43) .. (293.99,123.78) .. controls (295.66,122.13) and (297.33,122.14) .. (298.99,123.81) .. controls (300.64,125.49) and (302.31,125.5) .. (303.99,123.85) .. controls (305.66,122.2) and (307.33,122.21) .. (308.99,123.88) .. controls (310.65,125.55) and (312.32,125.56) .. (313.99,123.91) -- (317.49,123.94) -- (317.49,123.94) ;
	\draw [shift={(215.75,121.75)}, rotate = 180.39] [fill={rgb, 255:red, 0; green, 0; blue, 0 }  ][line width=0.08]  [draw opacity=0] (8.93,-4.29) -- (0,0) -- (8.93,4.29) -- cycle    ;
	
	\draw (80,131) node [anchor=north west][inner sep=0.75pt]   [align=left] {$\displaystyle \mu $};
	\draw (81,90) node [anchor=north west][inner sep=0.75pt]   [align=left] {$\displaystyle \nu $};
	\draw (319,130) node [anchor=north west][inner sep=0.75pt]   [align=left] {$\displaystyle \rho $};
	\draw (320,89) node [anchor=north west][inner sep=0.75pt]   [align=left] {$\displaystyle \sigma $};
	\draw (386,102) node [anchor=north west][inner sep=0.75pt]   [align=left] {$D_{\mu\nu,\rho\sigma}=\displaystyle \frac{i}{q^{2}} P_{\mu \nu ,\rho \sigma }$,};
	\draw (207,139) node [anchor=north west][inner sep=0.75pt]   [align=left] {$\displaystyle q$};

\end{tikzpicture}

where,
\begin{equation}
	P_{\mu\nu,\rho\sigma}=\frac 12(\eta_{\mu\rho}\eta_{\nu\sigma}+\eta_{\mu\sigma}\eta_{\nu\rho}-\eta_{\mu\nu}\eta_{\rho\sigma}).
\end{equation}
And
\tikzset{every picture/.style={line width=0.75pt}} 

\begin{tikzpicture}[x=0.75pt,y=0.75pt,yscale=-1,xscale=1]
	
	\draw    (104.01,119.5) .. controls (105.69,117.85) and (107.36,117.86) .. (109.01,119.53) .. controls (110.66,121.21) and (112.33,121.22) .. (114.01,119.57) .. controls (115.68,117.92) and (117.35,117.93) .. (119.01,119.6) .. controls (120.67,121.27) and (122.34,121.28) .. (124.01,119.63) .. controls (125.69,117.98) and (127.36,117.99) .. (129.01,119.67) .. controls (130.67,121.34) and (132.34,121.35) .. (134.01,119.7) .. controls (135.69,118.05) and (137.36,118.06) .. (139.01,119.74) .. controls (140.67,121.41) and (142.34,121.42) .. (144.01,119.77) .. controls (145.68,118.12) and (147.35,118.13) .. (149.01,119.8) .. controls (150.66,121.48) and (152.33,121.49) .. (154.01,119.84) .. controls (155.68,118.19) and (157.35,118.2) .. (159.01,119.87) .. controls (160.67,121.54) and (162.34,121.55) .. (164.01,119.9) .. controls (165.69,118.25) and (167.36,118.26) .. (169.01,119.94) .. controls (170.67,121.61) and (172.34,121.62) .. (174.01,119.97) .. controls (175.68,118.32) and (177.35,118.33) .. (179.01,120) .. controls (180.66,121.68) and (182.33,121.69) .. (184.01,120.04) .. controls (185.68,118.39) and (187.35,118.4) .. (189.01,120.07) .. controls (190.66,121.75) and (192.33,121.76) .. (194.01,120.11) .. controls (195.68,118.46) and (197.35,118.47) .. (199.01,120.14) .. controls (200.67,121.81) and (202.34,121.82) .. (204.01,120.17) .. controls (205.69,118.52) and (207.36,118.53) .. (209.01,120.21) .. controls (210.67,121.88) and (212.34,121.89) .. (214.01,120.24) .. controls (215.68,118.59) and (217.35,118.6) .. (219.01,120.27) .. controls (220.66,121.95) and (222.33,121.96) .. (224.01,120.31) .. controls (225.68,118.66) and (227.35,118.67) .. (229.01,120.34) .. controls (230.66,122.02) and (232.33,122.03) .. (234.01,120.38) .. controls (235.68,118.73) and (237.35,118.74) .. (239.01,120.41) .. controls (240.67,122.08) and (242.34,122.09) .. (244.01,120.44) .. controls (245.69,118.79) and (247.36,118.8) .. (249.01,120.48) .. controls (250.67,122.15) and (252.34,122.16) .. (254.01,120.51) .. controls (255.68,118.86) and (257.35,118.87) .. (259.01,120.54) .. controls (260.66,122.22) and (262.33,122.23) .. (264.01,120.58) .. controls (265.68,118.93) and (267.35,118.94) .. (269.01,120.61) .. controls (270.67,122.28) and (272.34,122.29) .. (274.01,120.64) .. controls (275.69,118.99) and (277.36,119) .. (279.01,120.68) .. controls (280.67,122.35) and (282.34,122.36) .. (284.01,120.71) .. controls (285.69,119.06) and (287.36,119.07) .. (289.01,120.75) .. controls (290.67,122.42) and (292.34,122.43) .. (294.01,120.78) .. controls (295.68,119.13) and (297.35,119.14) .. (299.01,120.81) .. controls (300.66,122.49) and (302.33,122.5) .. (304.01,120.85) .. controls (305.68,119.2) and (307.35,119.21) .. (309.01,120.88) .. controls (310.67,122.55) and (312.34,122.56) .. (314.01,120.91) -- (317.51,120.94) -- (317.51,120.94)(103.99,122.5) .. controls (105.66,120.85) and (107.33,120.86) .. (108.99,122.53) .. controls (110.64,124.21) and (112.31,124.22) .. (113.99,122.57) .. controls (115.66,120.92) and (117.33,120.93) .. (118.99,122.6) .. controls (120.65,124.27) and (122.32,124.28) .. (123.99,122.63) .. controls (125.67,120.98) and (127.34,120.99) .. (128.99,122.67) .. controls (130.65,124.34) and (132.32,124.35) .. (133.99,122.7) .. controls (135.67,121.05) and (137.34,121.06) .. (138.99,122.74) .. controls (140.65,124.41) and (142.32,124.42) .. (143.99,122.77) .. controls (145.66,121.12) and (147.33,121.13) .. (148.99,122.8) .. controls (150.64,124.48) and (152.31,124.49) .. (153.99,122.84) .. controls (155.66,121.19) and (157.33,121.2) .. (158.99,122.87) .. controls (160.65,124.54) and (162.32,124.55) .. (163.99,122.9) .. controls (165.67,121.25) and (167.34,121.26) .. (168.99,122.94) .. controls (170.65,124.61) and (172.32,124.62) .. (173.99,122.97) .. controls (175.66,121.32) and (177.33,121.33) .. (178.99,123) .. controls (180.64,124.68) and (182.31,124.69) .. (183.99,123.04) .. controls (185.66,121.39) and (187.33,121.4) .. (188.99,123.07) .. controls (190.64,124.75) and (192.31,124.76) .. (193.99,123.11) .. controls (195.66,121.46) and (197.33,121.47) .. (198.99,123.14) .. controls (200.65,124.81) and (202.32,124.82) .. (203.99,123.17) .. controls (205.67,121.52) and (207.34,121.53) .. (208.99,123.21) .. controls (210.65,124.88) and (212.32,124.89) .. (213.99,123.24) .. controls (215.66,121.59) and (217.33,121.6) .. (218.99,123.27) .. controls (220.64,124.95) and (222.31,124.96) .. (223.99,123.31) .. controls (225.66,121.66) and (227.33,121.67) .. (228.99,123.34) .. controls (230.64,125.02) and (232.31,125.03) .. (233.99,123.38) .. controls (235.66,121.73) and (237.33,121.74) .. (238.99,123.41) .. controls (240.65,125.08) and (242.32,125.09) .. (243.99,123.44) .. controls (245.67,121.79) and (247.34,121.8) .. (248.99,123.48) .. controls (250.65,125.15) and (252.32,125.16) .. (253.99,123.51) .. controls (255.66,121.86) and (257.33,121.87) .. (258.99,123.54) .. controls (260.64,125.22) and (262.31,125.23) .. (263.99,123.58) .. controls (265.66,121.93) and (267.33,121.94) .. (268.99,123.61) .. controls (270.65,125.28) and (272.32,125.29) .. (273.99,123.64) .. controls (275.67,121.99) and (277.34,122) .. (278.99,123.68) .. controls (280.65,125.35) and (282.32,125.36) .. (283.99,123.71) .. controls (285.67,122.06) and (287.34,122.07) .. (288.99,123.75) .. controls (290.65,125.42) and (292.32,125.43) .. (293.99,123.78) .. controls (295.66,122.13) and (297.33,122.14) .. (298.99,123.81) .. controls (300.64,125.49) and (302.31,125.5) .. (303.99,123.85) .. controls (305.66,122.2) and (307.33,122.21) .. (308.99,123.88) .. controls (310.65,125.55) and (312.32,125.56) .. (313.99,123.91) -- (317.49,123.94) -- (317.49,123.94) ;
	\draw [shift={(215.75,121.75)}, rotate = 180.39] [fill={rgb, 255:red, 0; green, 0; blue, 0 }  ][line width=0.08]  [draw opacity=0] (8.93,-4.29) -- (0,0) -- (8.93,4.29) -- cycle    ;
	\draw  [dash pattern={on 0.84pt off 2.51pt}]  (61.5,9.44) -- (104,121) ;
	\draw [shift={(80.44,59.14)}, rotate = 69.15] [fill={rgb, 255:red, 0; green, 0; blue, 0 }  ][line width=0.08]  [draw opacity=0] (8.93,-4.29) -- (0,0) -- (8.93,4.29) -- cycle    ;
	\draw  [dash pattern={on 0.84pt off 2.51pt}]  (104,121) -- (61.5,232.44) ;
	\draw [shift={(85.07,170.65)}, rotate = 110.88] [fill={rgb, 255:red, 0; green, 0; blue, 0 }  ][line width=0.08]  [draw opacity=0] (8.93,-4.29) -- (0,0) -- (8.93,4.29) -- cycle    ;

	\draw (331,137) node [anchor=north west][inner sep=0.75pt]   [align=left] {$\displaystyle \mu $};
	\draw (332,78) node [anchor=north west][inner sep=0.75pt]   [align=left] {$\displaystyle \nu $};
	\draw (207,139) node [anchor=north west][inner sep=0.75pt]   [align=left] {$\displaystyle q$};
	\draw (93,179) node [anchor=north west][inner sep=0.75pt]   [align=left] {$\displaystyle p$};
	\draw (100,49) node [anchor=north west][inner sep=0.75pt]   [align=left] {$\displaystyle p'$};
	\draw (393,104) node [anchor=north west][inner sep=0.75pt]   [align=left] {$\displaystyle V^\phi_{\mu\nu}=\frac{-i\kappa }2(p_{\mu } p'_{\nu } +p_{\nu } p'_{\mu } +\frac{1}{2} q^{2} \eta _{\mu \nu}) ,$};

\end{tikzpicture}

\tikzset{every picture/.style={line width=0.75pt}} 

\begin{tikzpicture}[x=0.75pt,y=0.75pt,yscale=-1,xscale=1]
	
	\draw    (75.5,61.44) -- (140,133) ;
	\draw [shift={(103.4,92.39)}, rotate = 47.97] [fill={rgb, 255:red, 0; green, 0; blue, 0 }  ][line width=0.08]  [draw opacity=0] (8.93,-4.29) -- (0,0) -- (8.93,4.29) -- cycle    ;
	\draw    (140,133) -- (74.5,225.44) ;
	\draw [shift={(111.01,173.92)}, rotate = 125.32] [fill={rgb, 255:red, 0; green, 0; blue, 0 }  ][line width=0.08]  [draw opacity=0] (8.93,-4.29) -- (0,0) -- (8.93,4.29) -- cycle    ;
	\draw    (139.99,131.5) .. controls (141.65,129.83) and (143.32,129.82) .. (144.99,131.48) .. controls (146.66,133.14) and (148.33,133.13) .. (149.99,131.46) .. controls (151.66,129.79) and (153.32,129.79) .. (154.99,131.45) .. controls (156.66,133.11) and (158.33,133.1) .. (159.99,131.43) .. controls (161.65,129.76) and (163.32,129.75) .. (164.99,131.41) .. controls (166.66,133.07) and (168.33,133.06) .. (169.99,131.39) .. controls (171.66,129.72) and (173.32,129.72) .. (174.99,131.38) .. controls (176.66,133.04) and (178.33,133.03) .. (179.99,131.36) .. controls (181.65,129.69) and (183.32,129.68) .. (184.99,131.34) .. controls (186.66,133) and (188.33,132.99) .. (189.99,131.32) .. controls (191.65,129.65) and (193.32,129.64) .. (194.99,131.3) .. controls (196.66,132.96) and (198.32,132.96) .. (199.99,131.29) .. controls (201.65,129.62) and (203.32,129.61) .. (204.99,131.27) .. controls (206.66,132.93) and (208.33,132.92) .. (209.99,131.25) .. controls (211.65,129.58) and (213.32,129.57) .. (214.99,131.23) .. controls (216.66,132.89) and (218.32,132.89) .. (219.99,131.22) .. controls (221.65,129.55) and (223.32,129.54) .. (224.99,131.2) .. controls (226.66,132.86) and (228.33,132.85) .. (229.99,131.18) .. controls (231.65,129.51) and (233.32,129.5) .. (234.99,131.16) .. controls (236.66,132.82) and (238.32,132.82) .. (239.99,131.15) .. controls (241.65,129.48) and (243.32,129.47) .. (244.99,131.13) .. controls (246.66,132.79) and (248.33,132.78) .. (249.99,131.11) .. controls (251.65,129.44) and (253.32,129.43) .. (254.99,131.09) .. controls (256.66,132.75) and (258.33,132.74) .. (259.99,131.07) .. controls (261.66,129.4) and (263.32,129.4) .. (264.99,131.06) .. controls (266.66,132.72) and (268.33,132.71) .. (269.99,131.04) .. controls (271.65,129.37) and (273.32,129.36) .. (274.99,131.02) .. controls (276.66,132.68) and (278.33,132.67) .. (279.99,131) .. controls (281.66,129.33) and (283.32,129.33) .. (284.99,130.99) .. controls (286.66,132.65) and (288.33,132.64) .. (289.99,130.97) .. controls (291.65,129.3) and (293.32,129.29) .. (294.99,130.95) -- (298.49,130.94) -- (298.49,130.94)(140.01,134.5) .. controls (141.67,132.83) and (143.34,132.82) .. (145.01,134.48) .. controls (146.68,136.14) and (148.35,136.13) .. (150.01,134.46) .. controls (151.68,132.79) and (153.34,132.79) .. (155.01,134.45) .. controls (156.68,136.11) and (158.35,136.1) .. (160.01,134.43) .. controls (161.67,132.76) and (163.34,132.75) .. (165.01,134.41) .. controls (166.68,136.07) and (168.35,136.06) .. (170.01,134.39) .. controls (171.68,132.72) and (173.34,132.72) .. (175.01,134.38) .. controls (176.68,136.04) and (178.35,136.03) .. (180.01,134.36) .. controls (181.67,132.69) and (183.34,132.68) .. (185.01,134.34) .. controls (186.68,136) and (188.35,135.99) .. (190.01,134.32) .. controls (191.66,132.65) and (193.33,132.64) .. (195,134.3) .. controls (196.67,135.96) and (198.33,135.96) .. (200,134.29) .. controls (201.66,132.62) and (203.33,132.61) .. (205,134.27) .. controls (206.67,135.93) and (208.34,135.92) .. (210,134.25) .. controls (211.66,132.58) and (213.33,132.57) .. (215,134.23) .. controls (216.67,135.89) and (218.33,135.89) .. (220,134.22) .. controls (221.66,132.55) and (223.33,132.54) .. (225,134.2) .. controls (226.67,135.86) and (228.34,135.85) .. (230,134.18) .. controls (231.66,132.51) and (233.33,132.5) .. (235,134.16) .. controls (236.67,135.82) and (238.33,135.82) .. (240,134.15) .. controls (241.66,132.48) and (243.33,132.47) .. (245,134.13) .. controls (246.67,135.79) and (248.34,135.78) .. (250,134.11) .. controls (251.66,132.44) and (253.33,132.43) .. (255,134.09) .. controls (256.67,135.75) and (258.34,135.74) .. (260,134.07) .. controls (261.67,132.4) and (263.33,132.4) .. (265,134.06) .. controls (266.67,135.72) and (268.34,135.71) .. (270,134.04) .. controls (271.66,132.37) and (273.33,132.36) .. (275,134.02) .. controls (276.67,135.68) and (278.34,135.67) .. (280,134) .. controls (281.67,132.33) and (283.33,132.33) .. (285,133.99) .. controls (286.67,135.65) and (288.34,135.64) .. (290,133.97) .. controls (291.66,132.3) and (293.33,132.29) .. (295,133.95) -- (298.51,133.94) -- (298.51,133.94) ;
	\draw [shift={(224.25,132.7)}, rotate = 179.8] [fill={rgb, 255:red, 0; green, 0; blue, 0 }  ][line width=0.08]  [draw opacity=0] (8.93,-4.29) -- (0,0) -- (8.93,4.29) -- cycle    ;

	\draw (127,184) node [anchor=north west][inner sep=0.75pt]   [align=left] {$\displaystyle p$};
	\draw (120,76) node [anchor=north west][inner sep=0.75pt]   [align=left] {$\displaystyle p'$};
	\draw (294,147) node [anchor=north west][inner sep=0.75pt]   [align=left] {$\displaystyle \mu $};
	\draw (295,103) node [anchor=north west][inner sep=0.75pt]   [align=left] {$\displaystyle \nu $};
	\draw (197,146) node [anchor=north west][inner sep=0.75pt]   [align=left] {$\displaystyle q$};
	\draw (322,111) node [anchor=north west][inner sep=0.75pt]   [align=left] {$\displaystyle V_{\mu \nu }^{\psi } =\frac{i\kappa }{8}[ 2\eta _{\mu \nu }( \slashed p +\slashed p' -2m)$\\$-( p+p')_{\mu } \gamma _{\nu } -\gamma _{\mu }( p +p')_{\nu }].$};

\end{tikzpicture}
\\
Here, the double-wavy line reprensents graviton, the dotted line reprensents scalar particle and the solid line represents Dirac particle.

We will calculate the differential scattering cross section of two scalar particles, a scalar particles and a Dirac particles, and two Dirac particles. And it is more convenient to perform calculations in the centre-of-mass frame. In this frame, four momenta of the particles are defined as follows:
\begin{equation}
	\left\{\begin{array}{l}
		p_1=\left(\sqrt{m_1^2+p_{\mathrm{cm}}^2}, 0,0, p_{\mathrm{cm}}\right) ,\\
		p_2=\left(\sqrt{m_2^2+p_{\mathrm{cm}}^2}, 0,0,-p_{\mathrm{cm}}\right), \\
		p_1^{\prime}=\left(\sqrt{m_1^2+p_{\mathrm{cm}}^2}, p_{\mathrm{cm}} \sin \theta, 0, p_{\mathrm{cm}} \cos \theta\right) ,\\
		p_2^{\prime}=\left(\sqrt{m_2^2+p_{\mathrm{cm}}^2},-p_{\mathrm{cm}} \sin \theta, 0,-p_{\mathrm{cm}} \cos \theta\right).
	\end{array}\right.
\end{equation}
Here, $p_1$ and $p_1^{\prime}$ are momenta of the first particle with mass $m_1$ before and after the scattering; $p_2$ and $p_2^{\prime}$ are momenta of the second particle with mass $m_2$ before and after the scattering; $p_{\mathrm{cm}}$ is three-momentum of particle in the centre of mass frame; $\theta$ is the centre of mass scattering angle. The following expressions give Mandelstam variables:
\begin{equation}
	\left\{\begin{array}{l}
		s=\left(p_1+p_2\right)^2=\left(\sqrt{m_1^2+p_{\mathrm{cm}}^2}+\sqrt{m_2^2+p_{\mathrm{cm}}^2}\right)^2 ,\\
		t=\left(p_1-p_1^{\prime}\right)^2=-4 p_{\mathrm{cm}}^2 \sin ^2 \frac{\theta}{2} ,\\
		u=\left(p_1-p_2^{\prime}\right)^2=\left(\sqrt{m_1^2+p_{\mathrm{cm}}^2}-\sqrt{m_2^2+p_{\mathrm{cm}}^2}\right)^2-4 p_{\mathrm{cm}}^2 \cos ^2 \frac{\theta}{2}.
	\end{array}\right.
\end{equation}

	\subsection{The differential scattering cross section of two scalar particles}
	\label{sect:3.1}
	
Gravitational scattering of two scalar particles has been discussed in Ref. \cite{Latosh23}. In this section, we will provide a brief overview of the results.

The matrix scattering element at tree level of two scalar particles is given by the following expression:
\begin{equation}
	\mathcal{M}=-i \frac{\kappa^2}{4 t}\left(s(s+t)-(2 s+t)\left(m_1^2+m_2^2\right)+m_1^4+m_2^4\right) .
\end{equation}
Using Eq. \eqref{1}, we obtain the result of the differential scattering cross section:
\begin{equation}
		\frac{\mathrm{d} \sigma_{ss}}{\mathrm{d} \Omega}=\frac{G^2}{4} \frac{1}{\sin ^4 \frac{\theta}{2}} \frac{\left(m_1^2 m_2^2+2 p_{\mathrm{cm}}^2\left(m_1^2+m_2^2\right)+4 p_{\mathrm{cm}}^2 \cos ^2 \frac{\theta}{2}\left(p_{\mathrm{cm}}^2+\sqrt{m_1^2+p_{\mathrm{cm}}^2} \sqrt{m_2^2+p_{\mathrm{cm}}^2}\right)\right)^2}{p_{\mathrm{cm}}^3\left(\sqrt{m_1^2+p_{\mathrm{cm}}^2}+\sqrt{m_2^2+p_{\mathrm{cm}}^2}\right) \sqrt{\left(p_{\mathrm{cm}}^2+\sqrt{m_1^2+p_{\mathrm{cm}}^2} \sqrt{m_2^2+p_{\mathrm{cm}}^2}\right)^2-m_1^2 m_2^2}} .
\end{equation}

First, we analyze the asymptotic behavior in the first case. Taking $p_{\mathrm{cm}}\rightarrow 0$, we obtain:
\begin{equation}
	\begin{aligned}
		\frac{\mathrm{d} \sigma_{ss}}{\mathrm{d} \Omega}= & \frac{1}{4}\left(G \mu m_1 m_2\right)^2 \frac{1}{p_{\mathrm{cm}}^4 \sin ^4 \frac{\theta}{2}}\left[1+p_{\mathrm{cm}}^2\left(\frac{4}{m_1^2}+\frac{4 \cos \theta+3}{m_1 m_2}+\frac{4}{m_2^2}\right)\right. \\
		& \left.+p_{\mathrm{cm}}^4\left(\frac{4}{m_1^4}+\frac{5(8 \cos \theta+5)}{4 m_1^3 m_2}+\frac{8 \cos ^2 \theta+16 \cos \theta+25}{2 m_1^2 m_2^2}+\frac{5(8 \cos \theta+5)}{4 m_1 m_2^3}+\frac{4}{m_2^4}\right)+\mathcal{O}\left(p_{\mathrm{cm}}^6\right)\right] ,
	\end{aligned}
\end{equation}
where, $\mu=(m_1^{-1}+m_2^{-1})^{-1}$. The leading term of this expression matches the result of classical Rutherford scattering, discussed in many textbooks \cite{Laudau76,Arnold97,Goldstein02}. Since both particles are scalar particles, we can make either particle as the target particle and the other particle as the scattered particle. If we take $m_1\gg m_2$, we obatain:
\begin{equation}{\label{13}}
	\begin{aligned}
		\frac{\mathrm{d} \sigma_{ss}}{\mathrm{d} \Omega}=  \frac{1}{4}\left(G  m_1 \right)^2 \frac{m_2^4}{p_{\mathrm{cm}}^4 \sin ^4 \frac{\theta}{2}}\left[1+\frac{4p^2_{\mathrm{cm}}}{m_2^2}+\frac{4p^4_{\mathrm{cm}}}{m_2^4}+\mathcal{O}\left(p_{\mathrm{cm}}^6\right)\right] .
	\end{aligned}
\end{equation}
We observe that the differential scattering cross section  $\mathrm{d}\sigma/\mathrm{d}\Omega$ can be expressed as: $\mathrm{d}\sigma/\mathrm{d}\Omega=f(m_1,v_2,\theta,\phi)$. This indicates that the result of differential scattering cross section is solely depends on the mass of the target particle and the incident velocity of the scattered particles, while it remains independent of the mass of the scattered particles. This result is consistent with the classical expression of WEP.

Next, we analyze the asymptotic behavior in the small angle limit in relativistic limit. Firstly, notations shall be modified: $m_1\rightarrow m$, $m_2\rightarrow M$, and $p_{\mathrm{cm}}\rightarrow\varepsilon m$. Secondly, one should take the leading order in $\theta\rightarrow 0$ and $m/M\rightarrow 0$:
\begin{equation}{\label{14}}
	\begin{aligned}
		\frac{d \sigma_{ss}}{d \Omega}= & \frac{(2 G M)^2}{\theta^4}\left[\left(2+\frac{1}{\varepsilon^2}\right)^2+2 \sqrt{1+\varepsilon^2}\left(4-\frac{1}{\varepsilon^4}\right) \frac{m}{M}+\mathcal{O}\left(\left(\frac{m}{M}\right)^2\right)\right] \\
		& +\frac{(2 G M)^2}{\theta^2}\left[\frac{1}{6}\left(2+\frac{1}{\varepsilon^2}\right)^2-\frac{1}{3} \sqrt{1+\varepsilon^2}\left(2+\frac{1}{\varepsilon^2}\right)\left(4+\frac{1}{\varepsilon^2}\right) \frac{m}{M}+\mathcal{O}\left(\left(\frac{m}{M}\right)^2\right)\right] \\
		& +(2 G M)^2\left[\frac{11}{720}\left(2+\frac{1}{\varepsilon^2}\right)^2-\frac{1}{360} \sqrt{1+\varepsilon^2}\left(76+\frac{60}{\varepsilon^2}+\frac{11}{\varepsilon^4}\right) \frac{m}{M}+\mathcal{O}\left(\left(\frac{m}{M}\right)^2\right)\right]+\mathcal{O}\left(\theta^2\right) .
	\end{aligned}
\end{equation}
We can find that the leading term of this expression matches the leading term of elastic Schwarzschild scattering cross
section \cite{Collins73}. Because the angular momentum carried by a Schwarzschild black hole is zero, we can regard it as a particle with a large mass and a spin of zero. Similarly, we can find that the result of differential scattering cross section does not depend on the mass of test particles. 
	
	\subsection{The differential scattering cross section of a scalar particle and a Dirac particle}
	{\label{sect:3.2}}
	
	The matrix scattering element at tree level of a scalar particle (particle 1) and a Dirac particle (particle 2) is given by the following expression:
	\begin{eqnarray}
		\mathcal{M}_{s_2,s'_2}
		&=&(p_{1\mu } p'_{1\nu } +p_{1\nu } p'_{1\mu } +\frac{1}{2} q^{2} \eta _{\mu \nu} )\frac{i\kappa^2}{32q^2}(\eta^{\mu\rho}\eta^{\nu\sigma}+\eta^{\mu\sigma}\eta^{\nu\rho}-\eta^{\mu\nu}\eta^{\rho\sigma})\nonumber\\
		& &\times \bar u^{s'_2}(p'_2)[2\eta _{\rho\sigma }( \slashed p_2 +\slashed p' _2-2m_2)-( p_2+p'_2)_{\rho} \gamma _{\sigma } -\gamma _{\rho }( p_2 +p_2')_{\sigma }]u^{s_2}(p_2)\nonumber\\
		&=&\frac{i\kappa^2}{32q^2}[(p^{\rho}_1 p'^{\sigma}_1 +p^{\sigma }_1 p'^{\rho}_2 +\frac{1}{2} q^{2} \eta ^{\rho \sigma})+(p^{\sigma}_1 p'^{\rho }_1 +p^{\rho }_1 p'^{\nu}_1 +\frac{1}{2} q^{2} \eta ^{\rho \sigma})]\nonumber\\
		& &\times \bar u^{s'_2}(p'_2)[2\eta _{\rho\sigma }( \slashed p_2 +\slashed p' _2-2m_2)-( p_2+p'_2)_{\rho} \gamma _{\sigma } -\gamma _{\rho }( p_2 +p_2')_{\sigma }]u^{s_2}(p_2)\nonumber\\
		& &-\frac{i\kappa^2}{32q^2}(2p_1\cdot p'_1+2q^2)\bar u^{s'_2}(p'_2)[8( \slashed p_2 +\slashed p' _2-2m_2)-2(\slashed p_2+\slashed p'_2)]u^{s_2}(p_2)\nonumber\\
		&=&\frac{i\kappa^2}{16q^2}\bar u^{s'_2}(p'_2)(Am_2+B(\slashed p_2+\slashed p'_2)+C\slashed p_1+D\slashed p'_1)u^{s_2}(p_2),
	\end{eqnarray}
	where,
	\begin{equation}
		A=4\left(2 m_1^2+t\right), \quad B=-2\left(m_1^2+t\right) ,\quad C=D=-\left(2 s+t-2 m_1^2-2 m_2^2\right).
	\end{equation}
Then, we can get the expression of the matrix element:
\begin{equation}{\label{17}}
	\mathcal M_{s_2,s'_2}=\frac{i\kappa^2}{16q^2}\bar u^{s'_2}(p'_2)(Am_2+B(\slashed p_2+\slashed p'_2)+C(\slashed p_1+\slashed p'_1))u^{s_2}(p_2).
\end{equation}
Here, we need to define the massive Dirac particles with a specific polarization direction \cite{Greiner03}:
\begin{equation}
	u^{s_2}\left(p_2\right)=\frac{1}{2}\left(1+\gamma_5 \slashed s_2\right) u\left(p_2\right),
\end{equation}
where,
\begin{equation}
	s^\mu_2=\left(\frac{\vec{p} \cdot \vec{s}_2|_{\text{rest}}}{m}, \vec{s}_2|_{\text{rest}}+\frac{\vec{p} \cdot \vec{s}_2|_{\text{rest}}}{m(m+E)} \vec{p}\right) .
\end{equation}
Here, $\vec s_2|_{\text{rest}}$ is polarization vector defined in the rest frame.

Using Eq. \eqref{2}, we obtain the result of $|M|^2$:
\begin{equation}
	\begin{aligned}
		|M|^2= & \left(\frac{\kappa^2}{16 t}\right)^2\left\{16 m_2^2 m_1^4\left(4 m_2^2-t\right)-\left[16 m_1^2 m_2^2-t\left(4 m_1^2-t\right)\right]\left(2 s+t-2 m_1^2-2 m_2^2\right)^2\right. \\
		& \left.+\left(2 s+t-2 m_1^2+2 m_2^2\right)^4\right\}.
	\end{aligned}
\end{equation}
We obtain the result of the differential scattering cross section:
\begin{equation}
	\begin{aligned}
		\frac{\mathrm{d} \sigma_{sd}}{\mathrm{d} \Omega}= & \frac{G^2}{256} \frac{1}{\sin ^4 \frac{\theta}{2}} \frac{1}{p_{\mathrm{cm}}^3\left(\sqrt{m_1^2+p_{\mathrm{cm}}^2}+\sqrt{m_2^2+p_{\mathrm{cm}}^2}\right) \sqrt{\left(p_{\mathrm{cm}}^2+\sqrt{m_1^2+p_{\mathrm{cm}}^2} \sqrt{m_2^2+p_{\mathrm{cm}}^2}\right)^2-m_1^2 m_2^2}} \\
		& \times\left\{16m_2^2 m_1^4\left(4 m_2^2+4 p_{\mathrm{cm}}^2 \sin ^2 \frac{\theta}{2}\right)-\left[16 m_1^2 m_2^2+4 p_{\mathrm{cm}}^2 \sin ^2 \frac{\theta}{2}\left(4 m_1^2+4 p_{\mathrm{cm}}^2 \sin ^2 \frac{\theta}{2}\right)\right]\right. \\
		& \left.\times\left(4 p_{\mathrm{cm}}^2 \cos ^2 \frac{\theta}{2}+4 \sqrt{m_1^2+p_{\mathrm{cm}}^2} \sqrt{m_2^2+p_{\mathrm{cm}}^2}\right)^2+\left(4 p_{\mathrm{cm}}^2 \cos ^2 \frac{\theta}{2}+4 \sqrt{m_1^2+p_{\mathrm{cm}}^2} \sqrt{m_2^2+p_{\mathrm{cm}}^2}\right)^4\right\}.
	\end{aligned}
\end{equation}

We follow the analytical method of asymptotic behavior in \ref{sect:3.1}. First, taking $p_{\mathrm{cm}}\rightarrow 0$, we obtain:
\begin{equation}
	\begin{aligned}
		\frac{d \sigma_{sd}}{d \Omega}= & \frac{1}{4}\left(G \mu m_1 m_2\right)^2 \frac{1}{p_{\mathrm{cm}}^4 \sin ^4 \frac{\theta}{2}}\left[1+p_{\mathrm{cm}}^2\left(\frac{4}{m_1^2}+\frac{4 \cos \theta+3}{m_1 m_2}+\frac{5+3 \cos \theta}{2 m_2^2}\right)\right. \\
		& +p_{\mathrm{cm}}^4\left(\frac{4}{m_1^4}+\frac{5(8 \cos \theta+5)}{4 m_1^3 m_2}+\frac{8 \cos ^2 \theta+20 \cos \theta+21}{2 m_1^2 m_2^2}+\frac{8 \cos ^2 \theta+34 \cos \theta+23}{4 m_1 m_2^3}\right. \\
		& \left.\left.+\frac{2+2 \cos \theta}{m_2^4}\right)+\mathcal{O}\left(p_{\mathrm{cm}}^6\right)\right].
	\end{aligned}
\end{equation}
The leading term of this expression also matches the result of classical Rutherford scattering. We first discuss the case where scalar particles are scattered particles and Dirac particles are target particles. If we take $m_1\ll m_2$, we obtain:
\begin{equation}{\label{23}}
	\begin{aligned}
		\frac{\mathrm{d} \sigma_{sd}}{\mathrm{d} \Omega}=  \frac{1}{4}\left(G  m_2 \right)^2 \frac{m_1^4}{p_{\mathrm{cm}}^4 \sin ^4 \frac{\theta}{2}}\left[1+\frac{4p^2_{\mathrm{cm}}}{m_1^2}+\frac{4p^4_{\mathrm{cm}}}{m_1^4}+\mathcal{O}\left(p_{\mathrm{cm}}^6\right)\right] .
	\end{aligned}
\end{equation}
This expression is the same as the differential scattering cross section of two scalar particles when target particle takes large mass limit expressed by Eq. \eqref{13}. Because the spin of scalar particles is zero, under the large mass limit of target particles, the differential scattering cross sections of scalar scattered particles and target particles do not depend on the spin properties of target particles. 

 Next, we discuss the case where Dirac particles are scattered particles and scalar particles are target particles. If we take $m_1\gg m_2$, we obtain:
 \begin{equation}{\label{24}}
 	\begin{aligned}
 		\frac{\mathrm{d} \sigma_{sd}}{\mathrm{d} \Omega}=  \frac{1}{4}\left(G  m_1 \right)^2 \frac{m_2^4}{p_{\mathrm{cm}}^4 \sin ^4 \frac{\theta}{2}}\left[1+\frac{(5+3\cos\theta)p^2_{\mathrm{cm}}}{2m_2^2}+\frac{(2+2\cos\theta)p^4_{\mathrm{cm}}}{m_2^4}+\mathcal{O}\left(p_{\mathrm{cm}}^6\right)\right] .
 	\end{aligned}
 \end{equation}
Compared with the differential scattering cross section of two scalar particles when target particle takes large mass limit expressed by Eq. \eqref{13}, it is found that under the large mass limit of target particles, the differential scattering cross sections of Dirac particles and scalar particles with target scalar particles have different angular distributions at the next-to-leading order. This means that the two-body scattering process through gravitational interaction can be used to distinguish the spin properties of matter, and the quantum WEP is broken. For non-relativistic scattering process, the difference is depressed by $ v^2$. The breaking of quantum WEP is difficult to find in the process of non-relativistic scattering. And for particles with the same spin property, the result of differential scattering cross section does not depend on the mass of scattered particles.

Next, we analyze the asymptotic behavior in the small angle limit in relativistic limit. If notations shall be modified: $m_1\rightarrow m$, $m_2\rightarrow M$, and $p_{\mathrm{cm}}\rightarrow\varepsilon m$, which means the scalar particles are scattered particles and Dirac particles are target particles, the expression is the same as the asymptotic behavior in the small angle limit of the differential scattering cross section of two scalar particles expressed by  Eq. \eqref{14}. We won't go into details here. Then notations shall be modified: $m_1\rightarrow M$, $m_2\rightarrow m$, and $p_{\mathrm{cm}}\rightarrow\varepsilon m$ , which means the Dirac particles are scattered particles and scalar particles are target particles. One should take the leading order in $\theta\rightarrow 0$ and $m/M\rightarrow 0$:
\begin{equation}
	\begin{aligned}
		\frac{d \sigma_{sd}}{d \Omega}= & \frac{(2 G M)^2}{\theta^4}\left[\left(2+\frac{1}{\varepsilon^2}\right)^2+2 \sqrt{1+\varepsilon^2}\left(4-\frac{1}{\varepsilon^4}\right) \frac{m}{M}+\mathcal{O}\left(\left(\frac{m}{M}\right)^2\right)\right] \\
		& +\frac{(2 G M)^2}{\theta^2}\left[\frac{1}{12}\left(-4-\frac{1}{\varepsilon^2}+\frac{2}{\varepsilon^4}\right)-\frac{1}{6} \sqrt{1+\varepsilon^2}\left(16+\frac{3}{\varepsilon^2}+\frac{2}{\varepsilon^4}\right) \frac{m}{M}+\mathcal{O}\left(\left(\frac{m}{M}\right)^2\right)\right] \\
		& +(2 G M)^2\left[\frac{1}{720}\left(-16-\frac{1}{\varepsilon^2}+\frac{11}{\varepsilon^4}\right)-\frac{1}{360} \sqrt{1+\varepsilon^2}\left(-104+\frac{15}{\varepsilon^2}+\frac{11}{\varepsilon^4}\right) \frac{m}{M}+\mathcal{O}\left(\left(\frac{m}{M}\right)^2\right)\right] \\
		& +\mathcal{O}\left(\theta^2\right).
	\end{aligned}
\end{equation}
Compared with the asymptotic behavior in the small angle limit of the differential scattering cross section of two scalar particles expressed by  Eq. \eqref{14}, it is found that under the mass limit of target particles, the differential scattering cross sections of Dirac particles and scalar particles with scalar target  particles have different angular distributions at  $\mathcal{O}(1/\theta^2)$. \newpage
	
	\subsection{The differential scattering cross section of two Dirac particles}
	
	The matrix scattering element of two Dirac particles is given by the following expression:
	\begin{equation}{\label{26}}
		\begin{aligned}
			\mathcal{M}_{s_1,s'_1,s_2,s'_2}= 
			& \frac{i \kappa}{8}\bar u^{s_1^{\prime}}\left(p_1^{\prime}\right) \left[2 \eta_{\mu \nu}\left(\slashed p_1+\slashed p_1^{\prime}-2 m_1\right)-\left(p_1+p_1^{\prime}\right)_\mu \gamma_\nu-\gamma_\mu\left(p_1+p_1^{\prime}\right)_\nu\right] u^{s_1}\left(p_1\right) \\
			& \times \frac{i}{2 q^2}\left(\eta^{\mu \rho} \eta^{\nu \sigma}+\eta^{\mu \sigma} \eta^{\nu \rho}-\eta^{\mu \nu} \eta^{\rho \sigma}\right) \\
			& \times \frac{i \kappa}{8} \bar{u}^{s^{\prime}_2}\left(p_2^{\prime}\right)\left[2 \eta_{\rho \sigma}\left(\slashed p_2+\slashed p_2^{\prime}-2 m_2\right)-\left(p_2+p_2^{\prime}\right)_\rho \gamma_\sigma-\gamma_\rho\left(p_2+p_2^{\prime}\right)_\sigma\right] u^{s_2}\left(p_2\right)\\
			=&-\frac{i \kappa^2}{128q^2}[\sum_{i=1}^4 c_i\bar u^{s_1^{\prime}}\left(p_1^{\prime}\right)\Gamma_i^1u^{s_1}\left(p_1\right) \bar u^{s_2^{\prime}}\left(p_2^{\prime}\right)\Gamma_i^2u^{s_2}\left(p_2\right) +A\bar u^{s_1^{\prime}}\left(p_1^{\prime}\right)\gamma^\mu u^{s_1}\left(p_1\right) \bar u^{s_2^{\prime}}\left(p_2^{\prime}\right)\gamma_\mu u^{s_2}\left(p_2\right)\\&+B^\nu_\mu \bar u^{s_1^{\prime}}\left(p_1^{\prime}\right)\gamma_\nu u^{s_1}\left(p_1\right) \bar u^{s_2^{\prime}}\left(p_2^{\prime}\right)\gamma^\mu u^{s_2}\left(p_2\right)],
		\end{aligned}
	\end{equation}
where,
\begin{equation}
	\begin{array}{ccc}
	c_1=-20,&\Gamma_1^1=\slashed p_1+\slashed p^{\prime}_1, &\Gamma_1^2=\slashed p_2+\slashed p^{\prime}_2,\\
	c_2=48,&\Gamma_2^1=m_1,&\Gamma_2^2=\slashed p_2+\slashed p^{\prime}_2,\\
	c_3=48,& \Gamma_3^1=\slashed p_1+\slashed p^{\prime}_1,& \Gamma_3^2=m_2,\\
	c_4=-128,&\Gamma_4^1=m_1,&\Gamma_4^2=m_2,
\end{array}
\end{equation}
and
\begin{equation}
	A=4[(p_1+p^{\prime}_1)\cdot(p_2+p_2^{\prime})],\quad B^\nu_\mu=4(p_1+p^{\prime}_1)_\mu(p_2+p_2^{\prime})^\nu.
\end{equation}

Using Eq. \eqref{2}, we obtain the result of $|M|^2$:
\begin{equation}
	\begin{aligned}
		|\mathcal{M}|^2
		= & \left(\frac{\kappa^2}{64 t}\right)^2\left\{8\left[\left(s-m_1^2-m_2^2\right)^2+\left(s+t-m_1^2-m_2^2\right)\right]\left(2 s+t-2 m_1^2-2 m_2^2\right)^2\right. \\
		& -4\left(4 m_1^2-t\right)\left(4 m_2^2-t\right)\left(2 s+t-2 m_1^2-2 m_2^2\right)^2+4\left(4 m_1^2-t\right)^2\left(4 m_2^2-t\right)^2 \\
		& -16t\left(2 s+t-2 m_1^2-2 m_2^2\right)^2\left( m_1^2+ m_2^2\right)+8t\left(4 m_1^2-t\right)\left(4 m_2^2-t\right) \\
		& \left.\times\left(4 m_1^2+4 m_2^2-2 t\right)+4 t^2\left(2 s+t-2 m_1^2-2 m_2^2\right)^2+12t^2\left(4 m_1^2-t\right)\left(4 m_2^2-t\right)\right\} .
	\end{aligned}
\end{equation}
Then we obtain the result of the differential scattering cross section:
\begin{equation}
	\begin{aligned}
		\frac{\mathrm{d} \sigma_{dd}}{\mathrm{d} \Omega}= & \frac{G^2}{4096} \frac{1}{\sin ^4 \frac{\theta}{2}} \frac{1}{p_{\mathrm{cm}}^3\left(\sqrt{m_1^2+p_{\mathrm{cm}}^2}+\sqrt{m_2^2+p_{\mathrm{cm}}^2}\right) \sqrt{\left(p_{\mathrm{cm}}^2+\sqrt{m_1^2+p_{\mathrm{cm}}^2} \sqrt{m_2^2+p_{\mathrm{cm}}^2}\right)^2-m_1^2 m_2^2}} \\
		& \times\left\{\left\{64\left[p_{\mathrm{cm}}^4+2 p_{\mathrm{cm}}^2 \sqrt{m_1^2+p_{\mathrm{cm}}^2} \sqrt{m_2^2+p_{\mathrm{cm}}^2}+\left(m_1^2+p_{\mathrm{cm}}^2\right)\left(m_2^2+p_{\mathrm{cm}}^2\right)\right]\right.\right. \\
		& \left.-128 p_{\mathrm{cm}}^2 \sin ^2 \frac{\theta}{2}\left(p_{\mathrm{cm}}^2+\sqrt{m_1^2+p_{\mathrm{cm}}^2} \sqrt{m_2^2+p_{\mathrm{cm}}^2}\right)+128 p_{\mathrm{cm}}^4 \sin ^4 \frac{\theta}{2}\right\}\\
		&\times\left(4 p_{\mathrm{cm}}^2 \cos ^2 \frac{\theta}{2}+4 \sqrt{m_1^2+p_{\mathrm{cm}}^2} \sqrt{m_2^2+p_{\mathrm{cm}}^2}\right)^2-4\left(4 m_1^2+4p_{\mathrm{cm}}^2 \sin^2 \frac{\theta}{2}\right)\left(4 m_2^2+4p_{\mathrm{cm}}^2 \sin ^2 \frac{\theta}{2}\right)\\&\times\left(4 p_{\mathrm{cm}}^2 \cos ^2 \frac{\theta}{2}+4 \sqrt{m_1^2+p_{\mathrm{cm}}^2} \sqrt{m_2^2+p_{\mathrm{cm}}^2}\right)^2+4\left(4 m_1^2+4p_{\mathrm{cm}}^2 \sin^2 \frac{\theta}{2}\right)^2\left(4 m_2^2+4p_{\mathrm{cm}}^2 \sin ^2 \frac{\theta}{2}\right)^2\\
		& +64p^2_{\mathrm{cm}}\sin^2\frac\theta2\left(4 p_{\mathrm{cm}}^2 \cos ^2 \frac{\theta}{2}+4 \sqrt{m_1^2+p_{\mathrm{cm}}^2} \sqrt{m_2^2+p_{\mathrm{cm}}^2}\right)^2\left( m_1^2+ m_2^2\right)\\
		&-32p^2_{\mathrm{cm}}\sin^2\frac\theta2\left(4 m_1^2+4p_{\mathrm{cm}}^2 \sin^2 \frac{\theta}{2}\right)\left(4 m_2^2+4p_{\mathrm{cm}}^2 \sin ^2 \frac{\theta}{2}\right)\left(4 m_1^2+4 m_2^2+8p^2_{\mathrm{cm}}\sin^2\frac\theta2\right)\\
		&+64p^4_{\mathrm{cm}}\sin^4\frac\theta2\left(4 p_{\mathrm{cm}}^2 \cos ^2 \frac{\theta}{2}+4 \sqrt{m_1^2+p_{\mathrm{cm}}^2} \sqrt{m_2^2+p_{\mathrm{cm}}^2}\right)^2+192p^4_{\mathrm{cm}}\sin^4\frac\theta2\\
		&\left.\times\left(4 m_1^2+4p_{\mathrm{cm}}^2 \sin^2 \frac{\theta}{2}\right)\left(4 m_2^2+4p_{\mathrm{cm}}^2 \sin ^2 \frac{\theta}{2}\right)\right\} .
	\end{aligned}
\end{equation}

We follow the analytical method of asymptotic behavior in \ref{sect:3.1}. First, taking $p_{\mathrm{cm}}\rightarrow 0$, we obtain:
\begin{equation}
	\begin{aligned}
		\frac{d \sigma_{dd}}{d \Omega}= & \frac{1}{4}\left(G \mu m_1 m_2\right)^2 \frac{1}{p_{\mathrm{cm}}^4 \sin ^4 \frac{\theta}{2}}\left[1+p_{\mathrm{cm}}^2\left(\frac{1}{m_1^2}+\frac{\cos \theta}{m_1 m_2}+\frac{1}{m_2^2}\right)\right. \\
		& +p_{\mathrm{cm}}^4\left(\frac{-\cos ^2 \theta+2 \cos \theta+3}{4 m_1^4}+\frac{10 \cos \theta+7}{4 m_1^3 m_2}+\frac{5 \cos ^2 \theta+8 \cos \theta+16}{2m_1^2 m_2^2}\right. \\
		& \left.\left.+\frac{10 \cos \theta+7}{4 m_1 m_2^3}+\frac{-\cos ^2 \theta+2 \cos \theta+3}{4m_2^4}\right)+\mathcal{O}\left(p_{\mathrm{cm}}^6\right)\right].
	\end{aligned}
\end{equation}
The leading order of this expression also matches the result of classical Rutherford scattering. Since both particles are Dirac particles, we can make either particle as the target particle and the other particle as the scattered particle. If we take $m_1\ll m_2$, we obtain:
\begin{equation}{\label{32}}
	\begin{aligned}
		\frac{\mathrm{d} \sigma_{dd}}{\mathrm{d} \Omega}=  \frac{1}{4}\left(G  m_2 \right)^2 \frac{m_1^4}{p_{\mathrm{cm}}^4 \sin ^4 \frac{\theta}{2}}\left[1+\frac{ p^2_{\mathrm{cm}}}{m_1^2}+\frac{(-\cos^2\theta+2\cos\theta+3)p^4_{\mathrm{cm}}}{4m_1^4}+\mathcal{O}\left(p_{\mathrm{cm}}^6\right)\right] .
	\end{aligned}
\end{equation}
Compared with the differential scattering cross section of  scalar scattered particles and Dirac target particles when target particle takes large mass limit expressed by Eq. \eqref{23}, it is found that under the large mass limit of  target particles, the differential scattering cross section of Dirac particles and scalar particles with  Dirac target particles have different angular distributions at the next-to-leading order. However, compared with the case of scalar target particles, if taking $\theta\rightarrow 0$, the differential scattering cross section of  Dirac scattered particles and scalar target particles when target particle takes large mass limit expressed by  Eq. \eqref{24} would be rewritten as
\begin{equation}
	\begin{aligned}
		\frac{\mathrm{d} \sigma_{sd}}{\mathrm{d} \Omega}=  \left(2G  m_1 \right)^2 \frac{m_2^4}{p_{\mathrm{cm}}^4 \theta ^4 }\left[1+\frac{4p^2_{\mathrm{cm}}}{m_2^2}+\frac{4p^4_{\mathrm{cm}}}{m_2^4}+\mathcal{O}\left(p_{\mathrm{cm}}^6\right)\right]+\mathcal{O}\left(\frac 1{\theta^2}\right) ,
	\end{aligned}
\end{equation}
which is the same as the differential scattering cross section of two scalar particles when target particle takes large mass limit expressed by Eq. \eqref{13} when taking $\theta\rightarrow 0$. And Eq. \eqref{32} would be rewritten as
\begin{equation}
	\begin{aligned}
		\frac{\mathrm{d} \sigma_{dd}}{\mathrm{d} \Omega}=  \left(2G  m_2 \right)^2 \frac{m_1^4}{p_{\mathrm{cm}}^4 \theta^4}\left[1+\frac{p^2_{\mathrm{cm}}}{m_1^2}+\frac{p^4_{\mathrm{cm}}}{m_1^4}+\mathcal{O}\left(p_{\mathrm{cm}}^6\right)\right]+\mathcal{O}\left(\frac 1{\theta^2}\right)  .
	\end{aligned}
\end{equation}
From this analysis, we find that the difference of differential scattering cross sections of light particles with different spin properties is at the next-to-leading order, whether heavy particles are scalar particles or Dirac particles. 

Next, we analyze the asymptotic behavior in the small angle limit in relativistic limit. Firstly, notations shall be modified: $m_1\rightarrow m$, $m_2\rightarrow M$, and $p_{\mathrm{cm}}\rightarrow\varepsilon m$. Secondly, one should take the leading order in $\theta\rightarrow 0$ and $m/M\rightarrow 0$:
\begin{equation}
	\begin{aligned}
		\frac{d \sigma_{dd}}{d \Omega}= & \frac{(2 G M)^2}{\theta^4}\left[\left(1+\frac{1}{\varepsilon^2}+\frac{1}{\varepsilon^4}\right)+2 \sqrt{1+\varepsilon^2}\left(1-\frac{1}{\varepsilon^4}\right) \frac{m}{M}+\mathcal{O}\left(\left(\frac{m}{M}\right)^2\right)\right] \\
		& +\frac{(2 G M)}{\theta^2}\left[\frac{1}{6}\left(1+\frac{1}{\varepsilon^2}+\frac{1}{\varepsilon^4}\right)-\frac{1}{6} \sqrt{1+\varepsilon^2}\left(4+\frac{3}{\varepsilon^2}+\frac{2}{\varepsilon^4}\right) \frac{m}{M}+\mathcal{O}\left(\left(\frac{m}{M}\right)^2\right)\right] \\
		& +(2 G M)^2\left[\frac{1}{720}\left(-34+\frac{11}{\varepsilon^2}+\frac{11}{\varepsilon^4}\right)-\frac{1}{360} \sqrt{1+\varepsilon^2}\left(-26+\frac{15}{\varepsilon^2}+\frac{11}{\varepsilon^4}\right) \frac{m}{M}+\mathcal{O}\left(\left(\frac{m}{M}\right)^2\right)\right] \\
		& +\mathcal{O}\left(\theta^2\right).
	\end{aligned}
\end{equation}
Compared with the asymptotic behavior in the small angle limit of the differential scattering cross section scalar scattered particles and Dirac target particles expressed by Eq. \eqref{14} (In Sect. \ref{sect:3.2}, we found that the asymptotic behavior in the small angle limit of the differential scattering cross section scalar scattered particles and Dirac target particles is the same as the asymptotic behavior in the small angle limit of the differential scattering cross section of two scalar particles expressed by  Eq. \eqref{14}),  it is found that under the large mass limit of target particles, the differential scattering cross sections of Dirac particles and scalar particles with Dirac target particles have different angular distributions at  $\mathcal{O}(1/\theta^4)$. 
	
	\section{The degree of polarization of scattered polarized particle }\label{sect:4}
	
	We can define the degree of polarization of scattered polarized particles by distinguishing the final polarization direction of light particles. It is defined as the difference between counting rates for positive and negative helicities, normalized to the total counting rate:
	\begin{equation}{\label{36}}
			P=\frac{\mathrm{d} \sigma\left(\lambda^{\prime}_{m}=+1\right)-\mathrm{d} \sigma\left(\lambda^{\prime}_{m}=-1\right)}{\mathrm{d} \sigma\left(\lambda^{\prime}_{m}=+1\right)+\mathrm{d} \sigma\left(\lambda^{\prime}_{m}=-1\right)}=\frac{1}{n_M}\sum_{s_M,s^{\prime}_M}\frac{|\mathcal{M}|^2(\lambda^{\prime}_{m}=1)-|\mathcal{M}|^2(\lambda^{\prime}_{m}=-1)}{|\mathcal{M}|^2_{NDFP}},
	\end{equation}
where, $\lambda^{\prime}_{m}$ marks the helicity of the final polarization. At this time, it is necessary to consider the angular relationship between the momentum direction and polarization direction of the initial and final states. In the centre of the mass frame, four momenta of the particles are defined as follows:
\begin{equation}
	\left\{\begin{array}{l}
		p_1=\left(\sqrt{m_1^2+p_{\mathrm{cm}}^2}, 0,0, p_{\mathrm{cm}}\right) ,\\
		p_2=\left(\sqrt{m_2^2+p_{\mathrm{cm}}^2}, 0,0,-p_{\mathrm{cm}}\right), \\
		p_1^{\prime}=\left(\sqrt{m_1^2+p_{\mathrm{cm}}^2}, p_{\mathrm{cm}} \sin \theta\cos\varphi, p_{\mathrm{cm}}\sin\theta\sin\varphi, p_{\mathrm{cm}} \cos \theta\right) ,\\
		p_2^{\prime}=\left(\sqrt{m_2^2+p_{\mathrm{cm}}^2},-p_{\mathrm{cm}} \sin \theta\cos\varphi, -p_{\mathrm{cm}}\sin\theta\sin\varphi,-p_{\mathrm{cm}} \cos \theta\right).
	\end{array}\right.
\end{equation}
The polarization direction of the initial state of polarized particles is in the left-handed coordinate system with the momentum direction of the initial state as the $z$-axis and in the rest frame is ${\vec s}|_{\text{rest}}=(\sin\alpha\cos\phi,\sin\alpha\sin\phi,\cos\alpha)$. The final polarization state of the probe polarized particles is helicity eigenstate, and its polarization direction is parallel or antiparallel to the momentum direction of the final state:
\begin{equation}
	\vec {s^{\prime}}=\lambda\frac{\vec {p^{\prime}}}{|\vec {p^{\prime}}|}, \quad \lambda=\pm 1.
\end{equation}

	\subsection{The degree of polarization  of scattered polarized particle with a scalar heavy particle}
	
	Using Eq. \eqref{17},  we obtain:
	\begin{equation}
		\begin{aligned}
			|\mathcal{M}(\lambda^{\prime}_{m_2}=1)|^2-	|\mathcal{M}(\lambda^{\prime}_{m_2}=-1)|^2= & \left(\frac{\kappa^2}{16 t}\right)^2\left\{32 m_2^2 m_1^4 f_1-16 m_1^2 m_2^2\left(2 s+t-2 m_1^2+2 m_2^2\right) f_2 \right.\\&\left.+4\left(2 s+t-2 m_1^2-2 m_2^2\right)^2f_3 \right\},
	\end{aligned}
	\end{equation}
	where,
	\begin{equation}
		\begin{aligned}
			f_1=&-p_{\mathrm{cm}}^2(1-\cos \theta) \cos \beta+2 m_2^2 \cos \theta \cos \alpha+2 m_2 \sqrt{p_{\mathrm{cm}}^2+m_2^2} \sin \theta \sin \alpha \cos (\varphi-\phi),\\
			f_2=&2 p_{\mathrm{cm}}^2(1+\cos \theta) \cos \alpha+4 \sqrt{p_{\mathrm{cm}}^2+m_1^2} \sqrt{p_{\mathrm{cm}}^2+m_2^2} \cos \theta \cos \alpha\\
			& +\left[\frac{2 p_{\mathrm{cm}}^2\left(\sqrt{p_{\mathrm{cm}}^2+m_1^2}+\sqrt{p_{\mathrm{cm}}^2+m_2^2}\right)}{m_2}+4 m_2 \sqrt{p_{\mathrm{cm}}^2+m_1^2}\right] \sin \theta \sin \alpha \cos (\varphi-\phi),\\
			f_3=&\left[4 m_1^2 m_2^2 \cos \theta+2 p_{\mathrm{cm}}^2 m_1^2(1+\cos \theta)+4 m_2^2 p_{\mathrm{cm}}^2 \cos \theta\right.\\
			& \left.+4 p_{\mathrm{cm}}^4(1+\cos \theta)+4 p_{\mathrm{cm}}^2 \sqrt{m_1^2+p_{\mathrm{cm}}^2} \sqrt{m_2^2+p_{\mathrm{cm}}^2}(1+\cos \theta)\right] \cos \alpha\\
			& +\left[4 m_2 p_{\mathrm{cm}}^2 \sqrt{p_{\mathrm{cm}}^2+m_1^2}+4 m_2 \sqrt{p_{\mathrm{cm}}^2+m_2^2}\left(p_{\mathrm{cm}}^2+m_1^2\right)\right] \sin \theta \sin \alpha\cos (\varphi-\phi).
		\end{aligned}
	\end{equation}
Next, using Eq. \eqref{36} and taking $p_{\mathrm{cm}}\rightarrow 0$, the expression of polarization degree can be obtained:
\begin{equation}
	\begin{aligned}
		P= & \left\{\cos \theta+p_{\mathrm{cm}}^2\left[\frac{-2 \cos ^2 \theta+2}{m_1 m_2}+\frac{-3 \cos ^2 \theta+3}{2 m_2^2}\right]+p_{\mathrm{cm}}^4\left[\frac{3 \cos ^2 \theta-3}{m_1^3 m_2}+\frac{5 \cos ^3 \theta+5 \cos ^2 \theta-5 \cos \theta-5}{m_1^2 m_2^2}\right. \right.\\
		& \left.\left.+\frac{7 \cos ^3 \theta+2 \cos ^2 \theta-7 \cos \theta-2}{m_1 m_2^3}+\frac{9 \cos ^3 \theta+7 \cos ^2 \theta-9 \cos \theta-7}{4 m_2^4}\right]+\mathcal{O}\left(p_{\mathrm{pm}}^6\right)\right\} \cos \alpha\\
		& +\left\{1-p_{\mathrm{cm}}^2\left[\frac{2 \cos \theta}{m_1 m_2}+\frac{3 \cos \theta}{2 m_2^2}\right]-\frac{p_{\mathrm{cm}}^4}{8}\left[\frac{24 \cos \theta}{m_1^3 m_2}+\frac{40 \cos ^2 \theta+14 \cos \theta-16}{m_1^2 m_2^2}\right.\right.\\
		&\left.\left.+\frac{56 \cos ^2 \theta+32 \cos \theta-24}{m_1 m_2^3}+\frac{18 \cos ^2 \theta+14 \cos \theta-9}{m_2^4}\right]+\mathcal{O}\left(p_{\mathrm{cm}}^6\right)\right\} \sin \theta \sin \alpha\cos (\varphi-\phi) .
	\end{aligned}
\end{equation}
Taking $m_1\gg m_2$, we can obtain
\begin{equation}
	\begin{aligned}
	P=&\cos\theta\cos\alpha+\sin\theta\sin\alpha\cos(\varphi-\phi)+p^2_{\mathrm{cm}}\left[\frac{-3 \cos ^2 \theta+3}{2 m_2^2}\cos\alpha-\frac{3 \cos \theta\sin\theta}{2 m_2^2}\sin\alpha\cos(\varphi-\phi)\right]\\
	&+p^4_{\mathrm{cm}}\left[\frac{9 \cos ^3 \theta+7 \cos ^2\theta-9 \cos \theta-7}{4 m_2^4}\cos\alpha-\frac{(18 \cos ^2 \theta+14 \cos \theta-9)\sin\theta}{8m_2^4}\sin\alpha\cos(\varphi-\phi)\right]\\
	&+\mathcal{O}(p^6_{\mathrm{cm}}).
\end{aligned}
\end{equation}
We find that the degree of polarization, as an observable, depends on the initial polarization direction. For the non-relativistic case, the leading order effect is the geometric overlap between the initial polarization direction and the final momentum direction.  
	\subsection{The degree of polarization of scattered polarized particle scattering with a Dirac heavy particle}
	
	Using Eq. \eqref{26}, we obtain:
	\begin{equation}
		\begin{aligned}
			&\frac{1}{n_{m_2}}\sum_{s_{m_2},s^{\prime}_{m_2}}|\mathcal{M}(\lambda^{\prime}_{m_1}=1)|^2-	|\mathcal{M}(\lambda^{\prime}_{m_1}=-1)|^2\\ =&(\frac{\kappa^2}{64t})^2\left\{-32(2s+t-2m_1^2-2m_2^2)tg_0+256m_1^2m_2^2(4m_2^2-t)g_1\right.\\
			&+(-512m_1^2m_2^2+64m_1^2t)(2s+t-2m_1^2-2m_2^2)g_2\\
			&\left.+[32(4m_1^2-t)t+64(2s+t-2m_1^2-2m_2^2)^2]g_3\right\}.
		\end{aligned}
	\end{equation}
where,
\begin{equation}
	\begin{aligned}
			g_0=&(p_{\mathrm{cm}}^2(1-\cos \theta)-m_1^2 \cos \theta)\cos \alpha-m_1\sqrt{p_{\mathrm{cm}}^2+m_1^2} \sin \theta \sin \alpha \cos (\phi-\psi),\\
			g_1=&-p_{\mathrm{cm}}^2(1-\cos \theta) \cos \beta+2 m_1^2 \cos \theta \cos \alpha+2 m_1 \sqrt{p_{\mathrm{cm}}^2+m_1^2} \sin \theta \sin \alpha \cos (\varphi-\phi),\\
			g_2=&2 p_{\mathrm{cm}}^2(1+\cos \theta) \cos \alpha+4 \sqrt{p_{\mathrm{cm}}^2+m_1^2} \sqrt{p_{\mathrm{cm}}^2+m_2^2} \cos \theta \cos \alpha\\
			& +\left[\frac{2 p_{\mathrm{cm}}^2\left(\sqrt{p_{\mathrm{cm}}^2+m_1^2}+\sqrt{p_{\mathrm{cm}}^2+m_2^2}\right)}{m_1}+4 m_1 \sqrt{p_{\mathrm{cm}}^2+m_2^2}\right] \sin \theta \sin \alpha \cos (\varphi-\phi),\\
			g_3=&\left[4 m_1^2 m_2^2 \cos \theta+2 p_{\mathrm{cm}}^2 m_2^2(1+\cos \theta)+4 m_1^2 p_{\mathrm{cm}}^2 \cos \theta\right.\\
			& \left.+4 p_{\mathrm{cm}}^4(1+\cos \theta)+4 p_{\mathrm{cm}}^2 \sqrt{m_1^2+p_{\mathrm{cm}}^2} \sqrt{m_2^2+p_{\mathrm{cm}}^2}(1+\cos \theta)\right] \cos \alpha\\
			& +\left[4 m_1 p_{\mathrm{cm}}^2 \sqrt{p_{\mathrm{cm}}^2+m_2^2}+4 m_1 \sqrt{p_{\mathrm{cm}}^2+m_1^2}\left(p_{\mathrm{cm}}^2+m_2^2\right)\right] \sin \theta \sin \alpha\cos (\varphi-\phi).
		\end{aligned}
	\end{equation}

Using Eq. \eqref{36} and taking $p_{\mathrm{cm}}\rightarrow 0$, the expression of polarization degree can be obtained:

	\begin{align}
		P= & \left\{\cos \theta+\frac{p_{\mathrm{cm}}^2}{4}\left[\frac{6 \cos \theta+6}{m_1^2}+\frac{4 \cos ^2 \theta+12 \cos \theta+8}{m_1 m_2}+\frac{-2\cos ^2 \theta+14\cos \theta}{m_2^2}\right]\right. \notag\\
		& +\frac{p_{\mathrm{cm}}^4}{4}\left[\frac{2\cos ^3 \theta-4 \cos ^2 \theta-4\cos\theta+2}{m_1^4}+\frac{-14 \cos ^2 \theta-4 \cos \theta+10}{m_1^3 m_2}\right. \notag\\
		& +\frac{-11 \cos ^3 \theta-14 \cos ^2 \theta+13 \cos \theta+28}{m_1^2 m_2^2}+\frac{2 \cos ^3 \theta-16\cos ^2 \theta-6 \cos \theta+12}{m_1 m_2^3} \notag\\
		& \left.\left.+\frac{2\cos ^3 \theta-2 \cos ^2 \theta-4 \cos \theta}{m_2^4}\right]+\mathcal{O}\left(p_{\mathrm{pm}}^6\right)\right\} \cos \alpha \notag\\
		& +\left\{1+\frac{p_{\mathrm{cm}}^2}{2}\left[\frac{3}{m_1^2}+\frac{2 \cos \theta+6}{m_1 m_2}+\frac{- \cos \theta+7}{m_2^2}\right]\right. \notag\\
		& +\frac{p_{\mathrm{cm}}^4}{8}\left[\frac{4\cos ^2 \theta-8 \cos \theta-9}{m_1^4}+\frac{-20 \cos \theta-20}{m_1^3 m_2}+\frac{-14 \cos ^2 \theta-14 \cos \theta+12}{m_1^2 m_2^2}\right. \notag\\
		& \left.\left.+\frac{4 \cos ^2 \theta-16 \cos \theta+14}{m_1 m_2^3}+\frac{4 \cos ^2 \theta-4 \cos \theta}{m_2^4}\right]+\mathcal{O}\left(p_{\mathrm{cm}}^6\right)\right\} \sin \theta \sin \alpha \cos (\varphi-\phi) .
	\end{align}
Then ,if taking $m_1\ll m_2$, we can obtain
\begin{equation}
	\begin{aligned}
		P=&\cos\theta\cos\alpha+\sin\theta\sin\alpha\cos(\varphi-\phi)+p^2_{\mathrm{cm}}\left[\frac{3\cos\theta+3}{2m_1^2}\cos\alpha\right.\\
		&\left.+\frac{3\sin\theta}{2 m_1^2}\sin\alpha\cos(\varphi-\phi)\right]+p^4_{\mathrm{cm}}\left[\frac{\cos ^3 \theta-2 \cos ^2\theta-2 \cos \theta+1}{2 m_1^4}\cos\alpha\right.\\
		&\left.+\frac{(4\cos ^2 \theta-8 \cos \theta-9)\sin\theta}{8m_1^4}\sin\alpha\cos(\varphi-\phi)\right]+\mathcal{O}(p^6_{\mathrm{cm}}).
	\end{aligned}
\end{equation}
We also find that the degree of polarization depends on the initial polarization direction, and for non-relativistic cases,  the leading order effect is the geometric overlap between the initial polarization direction and the final momentum direction.  

	\section{Summary and Outlook}
	\label{sect:5}
	
	In this paper, we calculated the differential scattering cross sections of two scalar particles, a scalar particle and a Dirac particle, and two Dirac particles at tree level. We found that when the mass of target particles is taking large mass limit, the angular distribution of differential scattering cross sections of particles with different spin properties of scattered particles is different. In other words, \textit{we can distinguish quantum matter by gravity with differential scattering cross section with the same target particles when taking large mass limit at tree level}. We analyzed the results of differential scattering cross section in non-relativistic and relativistic cases.
	
	When studying the asymptotic behavior  in non-relativistic case ($p_{\mathrm{cm}}\rightarrow 0$), we found that the leading effect of differential scattering cross sections of particles with different spin properties is consistent with the results of Rutherford scattering calculated by classical mechanics. The difference of differential scattering cross sections of particles with different spin properties is mainly reflected in the next-to-leading order. This difference is depressed by $(p _ {\mathrm {cm}}/m)^ 2$ in non-relativistic case. Therefore, the quantum WEP is weakly broken in the case of non-relativistic case.
	 
	 In relativistic case, we studied the asymptotic behavior of small angle limit ($\theta\rightarrow 0$). If the target particles are scalar particles, the differential scattering cross sections of Dirac particles and scalar particles have different angular distributions at $\mathcal{O}(1/\theta^2)$. And if the target particles are Dirac particles, the differential scattering cross sections of Dirac particles and scalar particles have different angular distributions at $\mathcal{O}(1/\theta^4)$. 
	 
	 We calculated the degree of polarization of the differential scattering cross section of light Dirac particles scattered by heavy scalar particles and heavy Dirac particles. We found that the leading order of polarization of differential scattering cross section $P\sim \cos\theta\cos\alpha+\sin\theta\sin\alpha\cos(\varphi-\phi)$. This result is the geometric overlap between the initial polarization direction and the final momentum direction. We take the degree of polarization of differential scattering cross section as a pointer, and the result depends on the polarization direction of incident particles.
	 
	 Taking the differential scattering cross section as a pointer, the reason why the quantum weak equivalence principle is broken is the nonlocality of quantum states. The trajectory of non-local classical objects is described by MPD equation, which predicts that the trajectories of classical objects with different rotational angular velocities are actually different when they move at the same initial velocity. For the quantum state, the trajectory is not a direct observable, and it often needs to be defined indirectly by some methods (such as EMT \cite{Lian22}). The differential scattering amplitude is a pointer that can be used to directly express the nonlocality of this quantum state breaking the quantum WEP.
	 
	 The particles carrying polarization are not only Dirac particles, but also Proca particles. In the future, we will calculate the differential scattering cross sections between Proca particles and scalar particles, Dirac particles and Proca particles through gravitational interaction.  In addition, we can also compare the differential scattering behavior of massless particles (photons and gravitons) scattered by heavy particles through gravitational interaction. For polarized particles, we can also calculate the degree of polarization of differential scattering cross sections of polarized particles with different polarization directions to observe the influence of different polarization directions on scattering results.
	 
	 We compared the particles with different spin characteristics scattered by scalar target particles with the results calculated by Ref. \cite{Accioly09}, and found that the results under the condition that the mass of target particles is infinite are consistent with the results that the gravitational field generated by a mass point is regarded as a classical external source. Then we can calculate the scattering cross section of polarized particles scattered by classical external sources applied by a mass point with angular momentum and to compare with two-particles scattering.

\end{document}